\documentclass{emulateapj}
\usepackage{apjfonts}

\newcommand{\etal}{et al.}

\newcommand{\mbh}{M_{\rm BH}}
\newcommand{\lhost}{L_{\rm h}}
\newcommand{\jhost}{j_{\rm h}}
\newcommand{\ldummy}{L_{0}}

\shorttitle{Black Hole and Host Mass Evolution}
\shortauthors{Hopkins \etal}
\slugcomment{Accepted to ApJ, July 25, 2006}
\begin{document}

\title{An Upper Limit to the Degree of Evolution Between Supermassive Black
Holes and their Host Galaxies}
\author{Philip F. Hopkins\altaffilmark{1}, 
Brant Robertson\altaffilmark{1}, 
Elisabeth Krause\altaffilmark{1}, 
Lars Hernquist\altaffilmark{1}, 
Thomas J. Cox\altaffilmark{1}
}
\altaffiltext{1}{Harvard-Smithsonian Center for Astrophysics, 
60 Garden Street, Cambridge, MA 02138}

\begin{abstract}

We describe a model-independent integral constraint which defines an upper
limit to the allowed degree of evolution in the ratio of black hole
(BH) mass to host galaxy luminosity or mass, as a function of
redshift.  Essentially, if the BH/host ratio is excessive at redshift
$z$, then it would imply that the total mass density in BHs above some
$\mbh^{\rm min}$ is larger at that redshift than at $z=0$, which is
impossible.  This argument requires no knowledge of host or BH
properties, only a lower limit to the observed luminosity density in
the brightest galaxies at some $z$.  We calculate this upper limit
from a variety of luminosity and mass functions in different bands
from redshifts $z=0-2$.  We show that it is consistent with passive
evolution of spheroid populations (with a fixed $M_{\rm BH}/M_{\rm
host}$ relation) in all cases, and provides tighter constraints than
have generally been obtained previously, ruling out at
$\gtrsim6\sigma$ observational and theoretical estimates suggesting
that $M_{\rm BH}/M_{\rm host}$ was significantly larger at high
redshifts than locally, although relatively weak 
(factor $\lesssim2$ by $z=2$) evolution is still allowed. 
We discuss a variety of possible ``loopholes''
or changes in the BH/host populations and correlations, and show that
they typically lower the upper limits and strengthen our conclusions.

\end{abstract}


\section{Introduction}
\label{sec:intro}

Recent discoveries of tight correlations between the masses of
supermassive black holes (BHs) in the centers of nearby galaxies and
either the luminosity \citep[e.g.,][]{KormendyRichstone95}, mass
\citep{Magorrian98}, or velocity dispersion \citep{FM00,Gebhardt00} of
their host spheroids demonstrate a fundamental link between the growth
of BHs and galaxy formation.  Determining the evolution of these
correlations with redshift is critical for informing analytical models
\citep[e.g.,][]{SR98} and simulations \citep{DSH05,Robertson05} which
follow the co-formation of BHs and bulges, as well as theories which
relate the evolution and statistics of BH formation and quasar
activity to galaxy mergers \citep[e.g.,][]{H06a,H06e} and to the
remnant spheroid population \citep{H06b}.  Likewise, the significance
of observations tracing the buildup of spheroid populations
\citep[e.g.,][]{Cowie96,Wolf03,Bell04,Bundy05b} and associations between spheroids in
formation, mergers, and quasar hosts \citep[e.g.,][]{H06d} depends on
understanding the evolution of BH/host correlations.

Efforts to directly infer these correlations at redshifts $z>0$ are,
however, difficult and limited by the small numbers of objects with
observable hosts.  Consequently, different groups have reached
seemingly contradictory conclusions, with O[{\small III}] velocity
dispersion \citep{Shields03} and BH clustering
\citep{AdelbergerSteidel05} measurements finding no evolution in these
correlations out to redshifts $z\sim4$, while CO velocity dispersion
\citep{Walter04,Shields05} and host $R$-band luminosity \citep{Peng06}
observations imply significant evolution (with BHs becoming over-massive
relative to their hosts) occurring at $z\gtrsim1$. Spectral template
fitting \citep{Treu04,Woo06} suggests strong evolution at much lower
redshifts $z=0.37$, and comparison of radio quasar and galaxy
populations \citep{McLure06} indicates evolution of the form $\propto
(1+z)^{2}$. In addition to the difficulty of identifying host
properties at these redshifts, the measurements generally rely upon
virial line-width relations \citep[e.g.,][and references
therein]{Kaspi05} to determine BH masses, which have not been tested
at high redshifts and require that the BH be visible as an optical,
broad-line AGN.

It is therefore of interest to develop additional constraints on the
evolution of BH/host correlations, especially ones which are
independent of systematics in host measurements and selection effects
in considering only AGN populations. For example, 
\citet{Merloni04} compare the integrated 
quasar luminosity and stellar mass densities, and infer a much weaker 
evolution of the form $\propto(1+z)^{0.5}$, but this is subject to systematic uncertainties 
in quasar radiative efficiencies and (as discussed below) in observational completeness 
and galaxy type segregation at low luminosities.  In this paper, we formulate and
apply a model-independent integral constraint which provides an upper
limit to evolution in these correlations as a function of redshift,
and show that it rules out various proposed forms of evolution at high
significance.  We adopt a $\Omega_{\rm M}=0.3$,
$\Omega_{\Lambda}=0.7$, $H_{0}=70\,{\rm km\,s^{-1}\,Mpc^{-1}}$
cosmology, but the choice of $H_{0}$ factors out in our analysis and
the limits depend only weakly on $\Omega_{\rm M}$ and
$\Omega_{\Lambda}$.

\section{The Integral Constraint}
\label{sec:deriv}

If there exists {\em some} mean relationship between BH mass ($\mbh$)
and host mass or observed luminosity in a (rest-frame) band
($\lhost$), we can parameterize it with the coefficient $\mu(z)$,
\begin{equation}
\langle{\mbh}\rangle=\mu(z)\,\lhost.
\label{eqn:correlation}
\end{equation}
For simplicity in what follows, we adopt a linear correlation, but the
argument can be generalized trivially for {\em any} functional dependence
$\mbh=\mu(z)\,f_{z}(\lhost)$. The term $\lhost$ can represent either total
or strictly bulge/spheroid luminosity.  Integrating the space density
of both sides of Equation~(\ref{eqn:correlation}) down to some minimum
$\mbh^{\rm min}=\mu(z)\,\lhost^{\rm min}$, i.e.\
\begin{equation}
\int_{\mbh^{\rm min}}^{\infty}\mbh\,{\rm d}n(\mbh)
=\mu(z)\,\int_{\lhost^{\rm min}}^{\infty}\lhost\,{\rm d}n(\lhost)
\end{equation}
 gives an equivalent relation 
between BH mass density and galaxy (or spheroid) luminosity density ($\jhost$), 
\begin{equation}
\rho_{\rm BH}(\mbh>\mbh^{\rm min}\,,\,z)=\mu(z)\,\jhost(\lhost>\mbh^{\rm min}/\mu(z)\,,\,z) \, ,
\end{equation}
with the $z=0$ equivalent 
$\rho_{\rm BH}(\mbh>\mbh^{\rm min}\,,\,0)=\mu(0)\,\jhost(\lhost>\mbh^{\rm min}/\mu(0)\,,\,0)$.
Note that scatter in the $\mbh-\lhost$ relationship will introduce an additional 
multiplicative factor here, but this will generally cancel in our analysis (discussed further below). 
We divide the relations to obtain 
\begin{eqnarray}
\nonumber\frac{\rho_{\rm BH}(\mbh>\mbh^{\rm min}\,,\,z)}{\rho_{\rm BH}(\mbh>\mbh^{\rm min}\,,\,0)}&=&
\frac{\mu(z)}{\mu(0)}\,
\frac{\jhost(\lhost>\mbh^{\rm min}/\mu(z)\,,\,z)}{\jhost(\lhost>\mbh^{\rm min}/\mu(0)\,,\,0)}\\
\nonumber&=&\Gamma(z)\,\frac{\jhost(\lhost>\ldummy/\Gamma(z)\,,\,z)}{\jhost(\lhost>\ldummy\,,\,0)}\\
&\leq& 1 \, ,
\label{eqn:biglimit}
\end{eqnarray}
where the second equality comes from substitution with the definitions 
$\Gamma(z)\equiv\mu(z)/\mu(0)$ and $\ldummy\equiv \mbh^{\rm min}/\mu(0)$. 
Note that this notation for $\Gamma(z)$ is used by \citet{Peng06}, but in their case refers 
to individual systems, rather than the evolution of the mean relation.
The third equality 
comes from the physical fact that it is not possible to destroy BH mass -- 
the total mass above a given $\mbh^{\rm min}$ can never decrease from $z$ to the present. 
This is true for {\em every} $\mbh^{\rm min}$ and corresponding $\ldummy$, 
which implies an upper limit to $\Gamma(z)$ for arbitrary $\ldummy$, 
\begin{equation}
\Gamma_{\rm max}(z)\,\frac{\jhost(\lhost>\ldummy/\Gamma_{\rm max}(z)\,,\,z)}
{\jhost(\lhost>\ldummy\,,\,z=0)}
=1,
\label{eqn:limit}
\end{equation}
where again the $\jhost$ are defined in the same {\em rest-frame} band. 
Since $\jhost(\lhost>x)$ is by definition a monotonic decreasing function of $x$, 
the left hand side of Equation~(\ref{eqn:limit}) is necessarily a monotonic increasing function of 
$\Gamma(z)$, and the upper limit for a given $\ldummy$ 
is well-defined. Any other functional dependence 
where BH mass increases with host luminosity or mass 
at fixed redshift will give a qualitatively identical constraint 
(it just will not lend itself to such convenient notation). 

\begin{figure}
    \centering
    \epsscale{1.2}
    \plotone{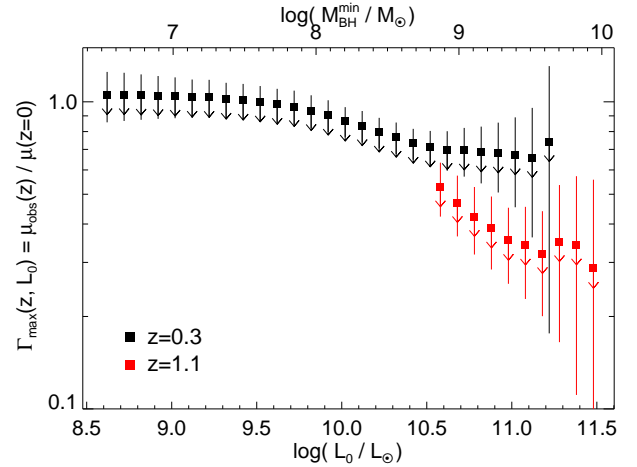}
    \caption{Upper limit to the allowed evolution in the proportionality between 
    BH mass and host (spheroid or total) $B$-band 
    luminosity ($\mu(z)/\mu(z=0)$, where 
    $M_{\rm BH}=\mu(z)\,\lhost$) as a function of the (arbitrary) minimum 
    luminosity ($\ldummy$) to which we integrate the luminosity density in 
    Equation~(\ref{eqn:limit}), at 
    $z=0.3$ and $z=1.1$ from comparison of the observed luminosity density in bright spheroids 
    \citep{Bell03,Faber05}. Points show the upper limits, error bars the uncertainty (dominated by 
    statistical uncertainty in the total luminosity density observed at $L>\ldummy$). 
    For comparison, the corresponding $\mbh^{\rm min}\equiv\mu(0)\,\ldummy$ is 
    also shown, although the actual values of $\mu(0)$ and $\ldummy$ are not important. 
    If $\Gamma(z)=\mu(z)/\mu(z=0)$ were greater than the limit shown at any $\ldummy$, then 
    the BH mass density above the corresponding $\mbh^{\rm min}$ would be larger at 
    $z$ than it is locally.
    Equation~(\ref{eqn:biglimit}) must hold for all $\ldummy$, so the lowest 
    statistically significant observational limit is the appropriate constraint. 
    \label{fig:demo}}
\end{figure}

Given a luminosity function (LF) or luminosity density above some 
luminosity $\jhost(\lhost>\ldummy)$ at $z=0$ and some higher redshift, 
we can determine the upper limit $\Gamma_{\rm max}(z)$ for 
every minimum luminosity $\ldummy$. 
Figure~\ref{fig:demo} illustrates how this depends on the ``dummy'' variable 
$\ldummy$, with an example of $\Gamma_{\rm max}(z,\,\ldummy)$ 
from Equation~(\ref{eqn:limit}). 
This is heuristic for now, but in detail we use the 
$g$-band observed local spheroid/red galaxy LF of \citet{Bell03} to 
calculate $\jhost(\lhost>\ldummy\,,\,z=0)$, and the $z=0.2-0.4$ and $z=1.0-1.2$ 
COMBO-17 $B$-band spheroid/red galaxy 
LFs from \citet{Faber05}
to calculate $\jhost(\lhost>\ldummy/\Gamma(z)\,,\,z)$ 
(see also \citet{Bell04,Willmer05} for details and 
the nearly negligible $B$-$g$ correction). We calculate the upper limit 
defined by Equation~\ref{eqn:limit}
($\Gamma_{\rm max}$) for every minimum luminosity 
$\ldummy$ at which the binned LF data exists (squares), 
with a $1\sigma$ upper limit (vertical error bar) 
estimated from the observational uncertainty in $\jhost(\lhost>\ldummy/\Gamma(z))$ 
using Equation~(\ref{eqn:limit})
to determine $\delta\,\Gamma_{\rm max}/\delta\jhost$
(technically we also add the errors in $\jhost$ from $z=0$ in quadrature, but this 
is in all cases dominated by the $\jhost(z>0)$ errors). 

How strong the constraint is depends on which $\ldummy$ we integrate
down to.  For example, integrating to low $\ldummy$ effectively
compares the cumulative BH mass densities at two redshifts, whereas
integrating to a relatively high $\ldummy$ compares the BH mass
density in only the most massive BHs. Both must be lower at $z$ than
locally, but if e.g.\ the majority of the most massive BHs are in
place by $z\sim1$ (such that their mass density is already comparable
to that observed locally) while only a small fraction of low mass BHs
exist (such that the total BH mass density may be substantially lower
than that at $z=0$), then the former integration will give a much
stricter limit $\Gamma_{\rm max}$. But since it is an upper limit
which must hold above {\em all} minimum masses or luminosities
$\ldummy$, the lowest value of $\Gamma_{\rm max}$ is the proper
constraint.

This tends to derive from the highest luminosities because
observations indicate that the stellar mass of the largest objects
is largely formed by $z\sim1$ \citep[``cosmic downsizing,''
e.g.,][]{Cowie96}, meaning that the luminosity density in them is
constant or rises with $z$, as the stellar populations are younger and
brighter at higher redshift.  As long as the number of objects at the
highest masses does not fall dramatically with $z$, this trend in the
constraints on $\Gamma(z)$ is preserved. In other words, if a
significant fraction of the most massive objects are formed by
$z\sim1$ and their rest-frame luminosity is typically brighter at that
time, then the BH mass to host luminosity ratio cannot also increase,
or the mass density in the most massive BHs would be much higher at
$z\sim1$ than it is today.

The observations are more complete and are dominated by spheroids
at these luminosities, which reduces our uncertainties and also means
that it does not matter whether the appropriate correlation is with
spheroid or total galaxy light/mass. However, at the brightest
luminosities, the number of observed objects is small and statistical
uncertainty is larger. Therefore, in what follows, we take as our
constraint the value of $\Gamma_{\rm max}$ which has the minimum
$1\sigma$ upper limit. This is a conservative assumption, as higher
$\ldummy$ will give lower (but less well-constrained) limits and the
cumulative significance of the upper limit (i.e.\ {\em cumulative}
$1\sigma$ upper error with respect to the entire data set) will always
be smaller than that we show.

\begin{figure*}
    \centering
    \epsscale{1.15}
    \plotone{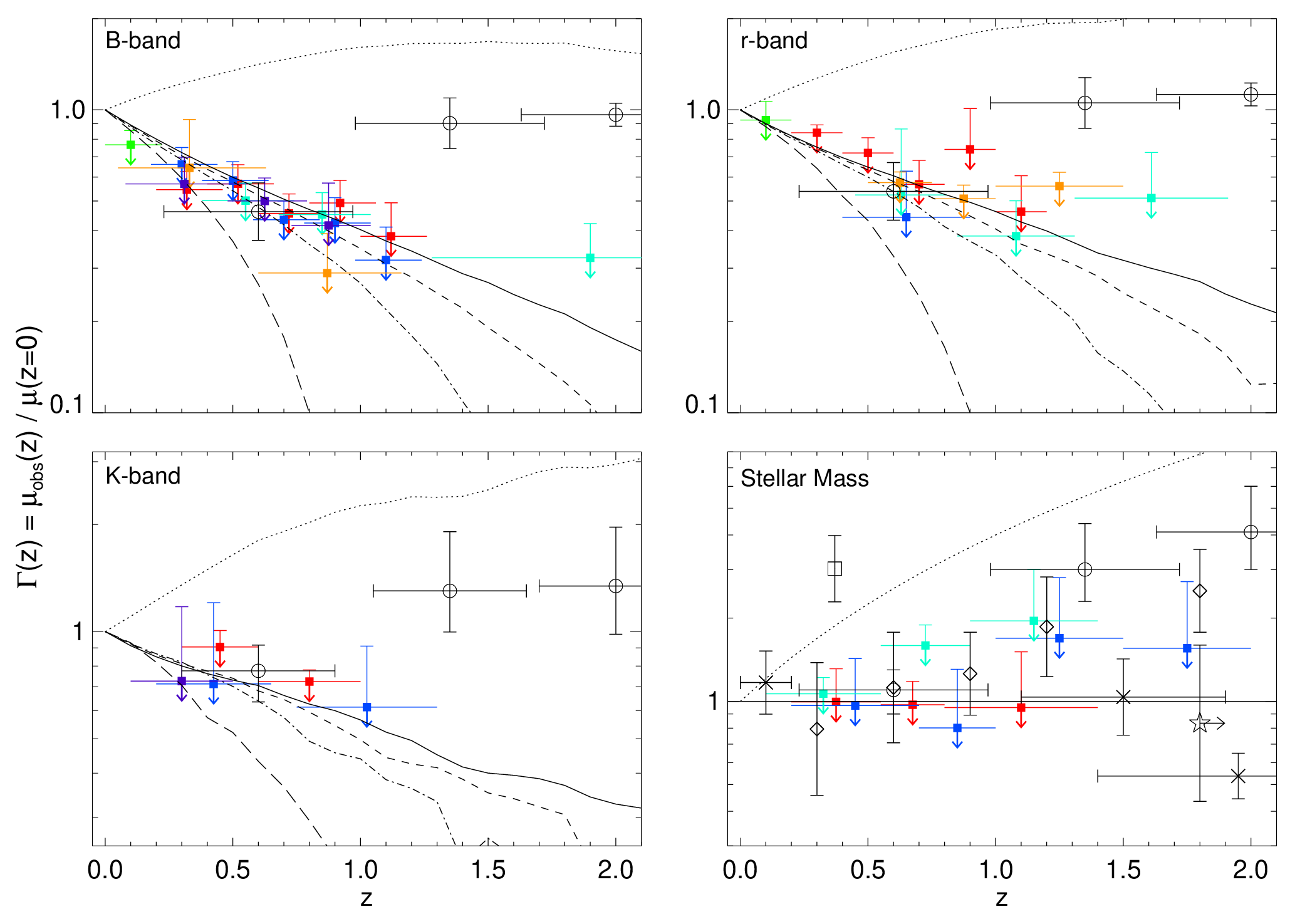}
    \caption{Upper limit to the allowed evolution in the proportionality between 
    BH and host (spheroid or total) 
    luminosity ($M_{\rm BH}=\mu(z)\,\lhost$) as a function of redshift, 
    based on the observed luminosity density or stellar mass density 
    as in Figure~\ref{fig:demo} (squares, with upper $1\sigma$ 
    error from observational uncertainties). 
    Limits in $B$-band are from \citet{Bell03} (SDSS+2MASS, local zero-point), 
    \citet{Blanton03} (SDSS, green), \citet{Im02} (DEEP GSS, orange), 
    \citet{Bell04,Willmer05,Faber05} (COMBO-17, blue, DEEP2, red), 
    \citet{Dahlen05} (GOODS, purple), \citet{Giallongo05} (COMBO-17, cyan). 
    In $r$-band from \citet{Nakamura03} (SDSS, local zero-point), \citet{Blanton03} (green), 
    \citet{Zucca05} (VVDS, blue), \citet{Gabasch06} (FDF, cyan), 
    \citet{Chen03} (LCIR Survey, orange), \citet{Wolf03} (COMBO-17, red). In $K$-band 
    from \citet{Bell03} (local zero-point), \citet{Cowie96} (Hawaii Deep Fields, red), 
    \citet{Pozzetti03} (K20, blue), \citet{Dahlen05} (purple). Mass functions are from 
    \citet{Bell03} (local zero-point), \citet{Bundy05a,Bundy05b} (GOODS+DEEP2, red), 
    \citet{Franceschini06} (IRAC/GOODS, cyan), \citet{Fontana04} (K20, blue). 
    All bands are rest-frame.
    Curves show the expectation of
    passive evolution tracks (with constant $M_{\rm BH}/M_{\ast}$), 
    with formation redshifts $z_{f}=1,\, 2,\, 3,\, 5$ (long-dashed, dot-dashed, 
    short-dashed, solid, respectively). 
    Open black points show the recent estimates of \citet[][crosses]{Shields03}, \citet[][square]{Woo06}, 
    \citet[][star]{AdelbergerSteidel05}, \citet[][circles]{Peng06}, \citet[][diamonds]{McLure06}; 
    dotted line shows $M_{\rm BH}\propto (1+z)^{2}\,M_{\ast}$ with $z_{f}=5$. The upper limits 
    are consistent with passive evolution in all cases, but strongly rule out a substantially larger 
    $\mbh/M_{\ast}$ at high redshift. 
    \label{fig:limits}}
\end{figure*}

With a well-defined $\Gamma_{\rm max}$ from any observed sample at a
given redshift, we now consider $\Gamma_{\rm max}(z)$.
Figure~\ref{fig:limits} shows this upper limit as a function of
redshift, from a large number of different observed LFs, in $B$, $r$,
and $K$ bands as well as estimated directly from stellar mass
functions. We calculate $\Gamma_{\rm max}$ (squares) and its error as
above in each case, and also plot (horizontal error bars) the redshift
range over which each binned LF is measured. Because it appears
locally that the appropriate correlation is between $\mbh$ and the
bulge/spheroid luminosity or mass, we consider only samples which have
measured LFs for both early and late types, and derive our constraints
only from the early type luminosity density comparison. We discuss
this below, however, and find it makes no difference to our
conclusions. For each band, we compare to the best-determined local
luminosity density as a function of the minimum luminosity
$\ldummy$. The resulting constraints are generally strongest from
optical LFs, because these typically involve the largest samples and
thus have the best statistics at the bright end. The mass-function
constraints, by comparison, are similar where the samples are large
enough to probe comparable number densities, but generally derive from
much smaller samples.

For comparison, we plot $\Gamma_{\rm max}(z)$ expected from passive 
evolution, assuming a constant BH to host stellar mass ratio ($\mbh/M_{\ast}$)
with the spheroid luminosity and colors passively evolving from 
a formation redshift $z_{f}=1,\, 2,\, 3,\, 5$ (long dashed, dotted, short dashed, solid lines, respectively) 
at solar metallicity according to the stellar population synthesis models of \citet{BC03}. 
We also plot (open black circles) the recent estimates of the evolution in these correlations from 
\citet{Peng06}. The authors directly measure the correlation with $R$; 
we use their adopted model to 
convert to mass, $B$, and $K$ band estimates. Note that they assume passive evolution 
with $z_{f}=5$; as they point out, allowing for star formation or younger populations 
will push their estimates of $\Gamma(z)$ to higher values. 
In the stellar mass panel, we also show (open black points) the 
observational estimates of \citet[][crosses]{Shields03}, \citet[][square]{Woo06}, 
\citet[][star]{AdelbergerSteidel05}, and \citet[][diamonds]{McLure06}. The 
dotted line in each panel 
shows $M_{\rm BH}\propto (1+z)^{2}\,M_{\ast}$ with $z_{f}=5$, a rough 
upper envelope of these observations as they would manifest in other bands.

\section{Discussion}
\label{sec:discuss}

Our model-independent integral measure provides a constraint on the
evolution in the correlation between BH mass and host luminosity or
mass.  The upper limits we derive are competitive with or stronger
than those from direct, but difficult observations of high-redshift
BHs and their hosts, and can be applied in any band (including those
for which direct estimates of these correlations at $z>0$ do not
exist).  Our upper limits are consistent with passive evolution (with
reasonable formation redshifts $z_{f}\lesssim6$) at all redshifts and
in every band, as well as hydrodynamic simulations of BH and spheroid
co-formation at different redshifts \citep{Robertson05}, and various
estimates of these correlations at low redshifts
\citep[$z\lesssim1$,][]{Shields03,Peng06}. The weak evolution 
inferred by \citet{Merloni04} cannot be ruled out for sufficiently 
low $z_{f}\lesssim3$, but the mass function data 
prefer a no-evolution case at $\sim2\,\sigma$.
However, the best-fit
evolution estimates (open points in Figure~\ref{fig:limits}) by e.g.\
\citet{Treu04,Shields05,Peng06,McLure06,Woo06} are ruled out at
$\gtrsim6\sigma$ with respect to the upper limits we measure over each
corresponding redshift interval. (Note that this does not necessarily
mean their measurements are inconsistent with our upper limits at
$6\sigma$, as the error bars in the direct observations are large and
generally inconsistent with passive evolution at only $\sim2-3\sigma$.) It should 
be noted, however, that there is a factor $\sim2$ systematic normalization uncertainty 
in the virial BH mass estimators these authors adopt, and lowering e.g.\ the 
estimates of \citet{Peng06} by such a factor (i.e.\ allowing for a 
factor of $\sim2$ increase in $\mbh/M_{\ast}$ by $z\sim2$) makes them consistent
with our upper limits. 

Given this apparent contradiction, it is natural to ask whether there
are effects which could increase the upper limits ($\Gamma_{\rm max}$)
in Figure~\ref{fig:limits}.  Observational incompleteness will only
lead to a higher ``true'' luminosity density at some $z$, and lower
$\Gamma_{\rm max}$ (assuming the luminosity density at $z=0$ is not
substantially less complete than that at higher redshift).  If we have
somehow severely overestimated the mass in large spheroids at high
$z$, i.e.\ if these red systems are really all dusty starbursts, this
could increase $\Gamma_{\rm max}$, but many of the samples shown in
Figure~\ref{fig:limits} are morphologically, not color, selected, and 
observations find that only a small fraction $\lesssim10\%$ of the
luminosity density in red galaxies at $z\sim1$ comes from dusty,
intrinsically blue sources (too small to change our conclusions)
\citep{Bell04b}. However, this may rise to $\sim30-50\%$ at $1\lesssim z\lesssim 2$
\citep[e.g.,][]{Cimatti02}, shifting the limits at these redshifts from the color-selected samples 
of \citet{Fontana04} and \citet{Giallongo05} up by a factor $\sim1.5-2$.

Furthermore, the most meaningful upper limits come from high
luminosities (typically $\sim0.5-1$\,mag above the LF break at each
$z$), where a significant fraction of the most massive galaxies are in
place by $z\sim1$ and this constraint is most sensitive to a change in
$\mu(z)$, but also where the observations are most well-constrained.
For this reason, it makes no difference which local LF we use to
compare in Equation~(\ref{eqn:limit}), as they all agree well at these
luminosities (see e.g.\ Figures~8 \& 9 of \citet{Bell03}).  We have
also directly tested this with the local LFs in
\citet{Kochanek01,Bell03,Blanton03,Nakamura03} and find the results
are similar.  Moreover, it makes no difference whether or not we use
the cumulative (all galaxy) LFs (i.e.\ assume a correlation between
$\mbh$ and total galaxy light or mass) instead of spheroid LFs as
above, since spheroids dominate the luminosity density above these
bright luminosities. In any case, the contribution of late-types tends
to evolve in a similar fashion to that of early types \citep[see e.g.\
the observations in Figure~\ref{fig:limits} and][]{Drory04}.  In even
more detail, allowing for late types but accounting for their bulges
(with more detailed morphology-dependent LFs) makes negligible
difference, as e.g.\ \citet{Koo05} find most massive bulges are
already in place at $z\sim1$, giving a similar constraint but with
more objects and thus a stronger $1\sigma$ upper limit.

Changing the form of the correlation also does not alter our results.
Our derivation of Eqn.~\ref{eqn:limit} can be performed for any
correlation of the form $\mbh=\mu(z)\,f(\lhost)$ or
$\mbh=f[\mu(z)\,\lhost]$, and any relationship where $\mbh$ is an
increasing function of host mass or luminosity will give a similar
result.  Furthermore, all non-linear forms of the correlation
$f(\lhost)$ which have been proposed observationally or theoretically
steepen at high luminosity (high $\mbh$), increasing the implied mass
of the most massive BHs and their relative weight in the BH mass
density, and directly considering these non-linear
\citep{KormendyRichstone95,MarconiHunt03,HaringRix04} or even
log-quadratic \citep{Wyithe06} correlations in every case {\em lowers}
$\Gamma_{\rm max}$.

But what if the shape or slope of $f(\lhost)$ changes with redshift?
This would be particularly interesting, 
in the implication that the ratio $\mbh/M_{\ast}$ evolves differently 
for systems of different mass. The exact effect on our derivation 
depends on the ``pivot point'' about which the relationship slope 
changes. For example, given that most of the 
$\mbh-\lhost$ measurements in Figure~\ref{fig:limits} have been 
restricted to high-$\mbh$, keeping these points 
fixed but finding a steeper slope at high redshift would imply a lower observed 
$\Gamma(z)$ at lower masses, which could relieve 
tension between our upper limits and the observations, although 
it would also {\em raise} the implied $\Gamma(z)$ at higher masses. 
In such a case, our integral constraint still obtains, it simply must 
be applied as a function of $\mbh^{\rm min}$ or $\ldummy$, as 
$\Gamma(z)$ becomes a function of mass as well as redshift. The 
limits shown in Figure~\ref{fig:limits} would then refer to the 
specific $\mbh^{\rm min}$ to which we perform our integration (generally 
$\sim10^{8}-10^{9}\,M_{\sun}$). 


Allowing for scatter in the BH/host relationship $f(\lhost)$ likewise
does not alter our results.  Non-zero scatter $\sigma(z)$ will
introduce a term $\sim\exp{\{0.5\,\ln^{2}{10}\,
(\sigma^{2}(z)-\sigma^{2}(0)) \}}$ in the left-hand side of
Equation~(\ref{eqn:limit}) (the appropriate exact factor will be a
non-trivial function of $\ldummy$, but since the relevant constraints
come from somewhat above the LF break, this factor is accurate to
$\sim10\%$ and the qualitative behavior is identical).  A larger
scatter implies more objects at the highest $\mbh$ for a given host
LF, so if the scatter in these relations were significantly smaller at
$z$, it would increase $\Gamma_{\rm max}$. However, every
observational indication, especially from the observations in conflict
with a low $\Gamma_{\rm max}$, is for an increasing scatter with $z$
\citep{Shields03,Shields05,Treu04,Peng06,Woo06}.  Moreover, if the
mean $\mbh$-host relation were higher in the past ($z\sim1$), we still
know observationally that there is a non-negligible population
(especially at the highest masses of interest) which has only
passively evolved since such times (no star formation beyond the
$\sim4\%$ level, and $M_{\rm BH}$ can only increase) \citep[see
e.g.,][]{Kauffmann03,Bell04,Faber05}, so these systems must have been
at or below the present relation at that redshift (spheroid-spheroid
``dry'' mergers, which by definition conserve the mean $\mbh$-host
mass/luminosity relations, cannot change this).  Therefore, the
scatter must increase with $z$ in such a scenario, which gives a {\em
lower} $\Gamma_{\rm max}$.

Since the limits on $\Gamma(z)$ rule out $\mbh$ increasing strongly in
the mean (i.e.\ for the general population) relative to the host bulge
or total galaxy, it is next worth asking whether the discrepant
$\Gamma(z)$ from some of the observations in Figure~\ref{fig:limits}
could represent some particular (short-lived) stage of
development. The simplest possibility is that once an object has
formed its bulge or spheroid, it will lie on the appropriate $z=0$
(passively evolved) relation, but that the observations see disks
which have grown their BHs but have not yet grown a bulge. In such a
scenario, $\Gamma(z)$ in spheroids is given by the passive evolution
tracks, and there is an independent, potentially much larger
$\Gamma(z)$ for ``pre-spheroids.'' 

This pre-spheroid population, however, cannot constitute all disks or
blue ``star-forming'' galaxies.  First, at least some disks have
survived since $z\sim1$, and this would demand they have overmassive
BHs relative to the total galaxy light/mass, where in fact the
opposite is observed \citep[i.e.\ $\mbh$ correlates with bulge, not
total galaxy mass; e.g.,][]{KormendyRichstone95}.  Second, this
scenario trivially modifies our calculation of $\Gamma_{\rm max}$ --
we integrate over spheroids with $\Gamma(z)$ from the passive
evolution tracks and determine the upper limit to $\Gamma(z)$ in the
remaining disk population. In Figure~\ref{fig:limits} it is clear that
the previous $\Gamma_{\rm max}$ is already marginally consistent with
the passive evolution tracks; adding an additional source of high-mass
BHs is not possible. Regardless, integrating down to a lower $\ldummy$
(where disks begin to dominate the population) gives a similar
constraint to those in Figure~\ref{fig:limits}, since there is little
room for ``additional'' BH mass density and, as noted above, the disk
population evolves similarly to spheroids at the bright end (in the
sense that the evolution is weak).

The remaining possibility is that objects with BHs larger than our
$\Gamma_{\rm max}$ upper limits are an even more limited sub-class of
galaxies. The natural expectation might be that AGN or bright quasars
in particular obey this discrepant relation, since essentially all of
the estimates of the high-redshift relation are from quasars. Locally
at least, bright quasars constitute a tiny fraction of galaxies, so
they can have much higher BH masses than their inactive 
counterparts (i.e.\ higher $\mu_{\rm QSO}/\mu(z=0)$) without
contributing much total BH mass. However, this phase would have to be
short-lived, much like the quasar phase itself, or else it is
equivalent to saying that a large fraction of non-spheroids obey this
higher $\mu(z)$ and violates the integral constraints as above. Therefore, 
the host must ``catch up'' to have the appropriate stellar mass for
the $z=0$ (passively evolved) relation in a short period of time. However,
cosmological cooling and infall generally requires a non-negligible
fraction of a Hubble time, implying (if this were the mechanism to add
bulge mass) that too large a fraction of galaxies would lie on the
discrepant relation (given the integral constraints above), and
invoking subsequent mergers or interactions to transport gas would
imply that the quasar somehow ``knows'' the system is going to
interact before the interaction begins.

This suggests that the mass required for the bulge to ``catch up'' is
present, in the form of gas ``about'' to be turned into stars \citep[see also
e.g.][]{Croton05}, but this hypothesis encounters a number of
observational conflicts. To reach the appropriate $z=0$ relation given 
the observations in Figure~\ref{fig:limits}
which imply strong evolution in $\mbh/M_{\ast}$, the
systems would have to be $\gtrsim75\%$ gas at $z\gtrsim1$, which for
the appropriate total mass range at these redshifts with these
concentrations gives quiescent (lower-limit) star formation rates
$\sim10-100$'s of $M_{\sun}\,{\rm yr^{-1}}$
\citep[e.g.,][]{Kennicutt98,SDH05b}. The typical quasar hosts would be
late-type galaxies, and in the subsequent, brief period of rapid star
formation the BH would be overmassive with the galaxy ``catching up''
(and could not significantly accrete without preventing this
``catching up''). 

All of these predictions appear to be opposite the observations, with
quasar spectra dominated by {\em post}-starburst activity
\citep{VandenBerk06}, typically early-type, bulge-dominated hosts
\citep[especially in the brightest quasars;][]{Dunlop03,Floyd04}, and
observed starbursts usually showing high accretion rate BHs which
are at that moment undermassive relative to their hosts
\citep[e.g.,][]{Borys05}.  Moreover, if star formation is ongoing
in these objects to grow their bulges, it implies a lower
mass-to-light ratio and suggests that the discrepancy between their
$\mbh$-host relations is even larger than that in
Figure~\ref{fig:limits}, amplifying these difficulties.  Furthermore,
by the highest luminosities and redshifts $z\sim1$, the active
fraction of BHs with masses $M_{\rm BH}\gtrsim10^{9}\,M_{\sun}$ is
$\gtrsim10\%$ \citep[relative to the total number of such objects at
$z\sim0$;][]{MD04}, which creates some tension if these are a
different population from the ``finished'' large spheroids at that
redshift, which can already account for the $z=0$ mass density of such
large BHs (but is perfectly consistent if these are the same
population, and quasars have large bulges).  Finally, associating this
discrepant $\Gamma(z)$ with quasars connects it not with the
cosmological buildup of galaxies but with a particular evolutionary
``snapshot'' or event. This makes it difficult to explain why quasars
at $z\gtrsim1$ would appear in bulge-less systems as a ``signpost'' of
bulge formation about to occur, while quasars of the same luminosity
at $z\lesssim1$ and low-luminosity AGN (which form a continuum in
$L_{\rm QSO}$) would appear in normal, massive, bulge-dominated
systems.

\acknowledgments We thank Chien Peng for very helpful discussion and 
clarifications. This work was supported in part by NSF grant AST
03-07690, and NASA ATP grants NAG5-12140, NAG5-13292, and NAG5-13381.

\end{document}